\documentclass[conference]{IEEEtran}
\IEEEoverridecommandlockouts
\usepackage{cite}
\usepackage{amsmath,amssymb,amsfonts}
\usepackage{algorithmic}
\usepackage{graphicx}
\usepackage{textcomp}
\usepackage{multirow}
\usepackage[table]{xcolor}
\usepackage{xcolor}
\usepackage[hidelinks]{hyperref}
\def\BibTeX{{\rm B\kern-.05em{\sc i\kern-.025em b}\kern-.08em
    T\kern-.1667em\lower.7ex\hbox{E}\kern-.125emX}}

\begin{document}

\title{SELF-CARE: \underline{Sel}ective \underline{F}usion with \underline{C}ontext-\underline{A}wa\underline{re} Low-Power Edge Computing for Stress Detection\\
\thanks{$^*$Both authors contributed equally to this research.}
}

\author{\IEEEauthorblockN{Nafiul Rashid$^{*1}$, Trier Mortlock$^{*2}$, Mohammad Abdullah Al Faruque$^{1,2}$}
		\IEEEauthorblockA{\textit{$^1$Department of Electrical Engineering and Computer Science}\\
			\textit{$^2$Department of Mechanical and Aerospace Engineering} \\
			\textit{University of California, Irvine, California, United States }\\
			\{nafiulr, tmortloc, alfaruqu\}@uci.edu}
}

\maketitle

\begin{abstract}

Detecting human stress levels and emotional states with physiological body-worn sensors is a complex task, but one with many health-related benefits.  
Robustness to sensor measurement noise and energy efficiency of low-power devices remain key challenges in stress detection. 
We propose \textbf{SELF-CARE}, a fully wrist-based method for stress detection that employs context-aware selective sensor fusion that dynamically adapts based on data from the sensors. Our method uses motion to determine the context of the system and learns to adjust the fused sensors accordingly, improving performance while maintaining energy efficiency. 
\textbf{SELF-CARE} obtains state-of-the-art performance across the publicly available WESAD dataset, achieving 86.34\% and 94.12\% accuracy for the 3-class and 2-class classification problems, respectively. Evaluation on real hardware shows that our approach achieves up to 2.2$\times$ (3-class) and 2.7$\times$ (2-class) energy efficiency compared to traditional sensor fusion. 

\end{abstract}


\section{Introduction}
\label{sec:intro}



The future of smart healthcare requires dependable sensor systems that can operate at increased levels of autonomy under energy constraints while providing valuable health-related information. 
One area gaining significant attention is\textit{ affective computing}, or the ability for machines to understand human emotional states. Stress detection is one example of affective computing that allows machines to detect stress levels within humans, which has a myriad of implications for healthcare science \cite{american2020stress}. 
Stress can be interpreted as a physiological state that is triggered by chemical or hormonal surges during moments of physical, cognitive, emotional, or acute challenges \cite{goldstein2010adrenal}.


Stress detection via physiological sensor data has been widely studied \cite{healey2005detecting,picard2001toward,koelstra2011deap,schmidt2018introducing}.
Physics-based models cannot relate this sensor data to explicit stress states, so classical machine learning models (random forests, decision trees, etc.) or deep learning models (convolutional neural networks, long short-term memory, etc.) are often used to perform stress classification as the models learn over labeled datasets \cite{samyoun2020stress, huynh2021stressnas, rashid2021feature, ragav2019scalable}.
However, classical machine learning models are more commonly used than deep learning approaches due to computational complexity and explainability \cite{schmidt2019multi,bota2020emotion}.



Methods using sensor fusion across multi-modal physiological data have been commonly used to increase performance of emotion recognition \cite{bota2020emotion}. \textit{Early}, or feature-level, fusion focuses on combining data at the raw-data level compared to \textit{late}, or decision-level, fusion that combines the final outputs of a system. Even methods that employ both early and late fusion are noticeably limited, since they have static architectures that cannot adapt to changing contexts \cite{malawade2022hydrafusion}. Another key challenge in using these physiological signals is that they are susceptible to large amounts of noise during physical motion. Fusing such noisy measurements can subsequently degrade the classification performance \cite{schmidt2019multi}. 
Lastly, there is a lack of focus in stress detection to evaluate feasibility for edge (on-device) computing \cite{rashid2022ahar} as solutions should be energy-efficient and capable of running on resource-constrained devices \cite{malawade2022ecofusion}.

Key research challenges arise from the current methods for stress detection, notably: (i) how to develop an adaptive architecture that alters the fusion schema depending on the current context; (ii) how to utilize measurements from the motion sensors to model the context; (iii) how to achieve comparable results to higher fidelity chest-worn devices while using more energy-efficient wrist wearable sensors; and (iv) how to incorporate temporal aspects into the stress classification problem that can further improve accuracy.



To address these challenges, we propose \textsc{SELF-CARE}, a fully wrist-based solution that models context as a function of motion and proposes a selective sensor fusion that adapts based on this learned context. SELF-CARE outperforms existing solutions for stress detection, while also providing energy efficiency suitable for edge computing on the wrist. 
We present the following key contributions:
\begin {enumerate}
\item We introduce a selective sensor fusion method that learns context based on motion, and dynamically adjusts the sensor fusion performed to maximize classification performance while ensuring energy efficiency.
\item We propose a late fusion technique for classification using a Kalman filter that incorporates temporal dynamics.
\item We validate our methodology on the WESAD dataset, showing that SELF-CARE achieves state-of-the-art performance for the 3-class and 2-class stress detection problems while using only wrist-worn sensors.
\item Experimental evaluation on real hardware shows that our SELF-CARE methodology is feasible for on-device, energy-efficient computing.

\end{enumerate}

\section{Methodology}
\label{sec:meth}
\begin{figure}
    \centering
    \includegraphics[width=\linewidth]{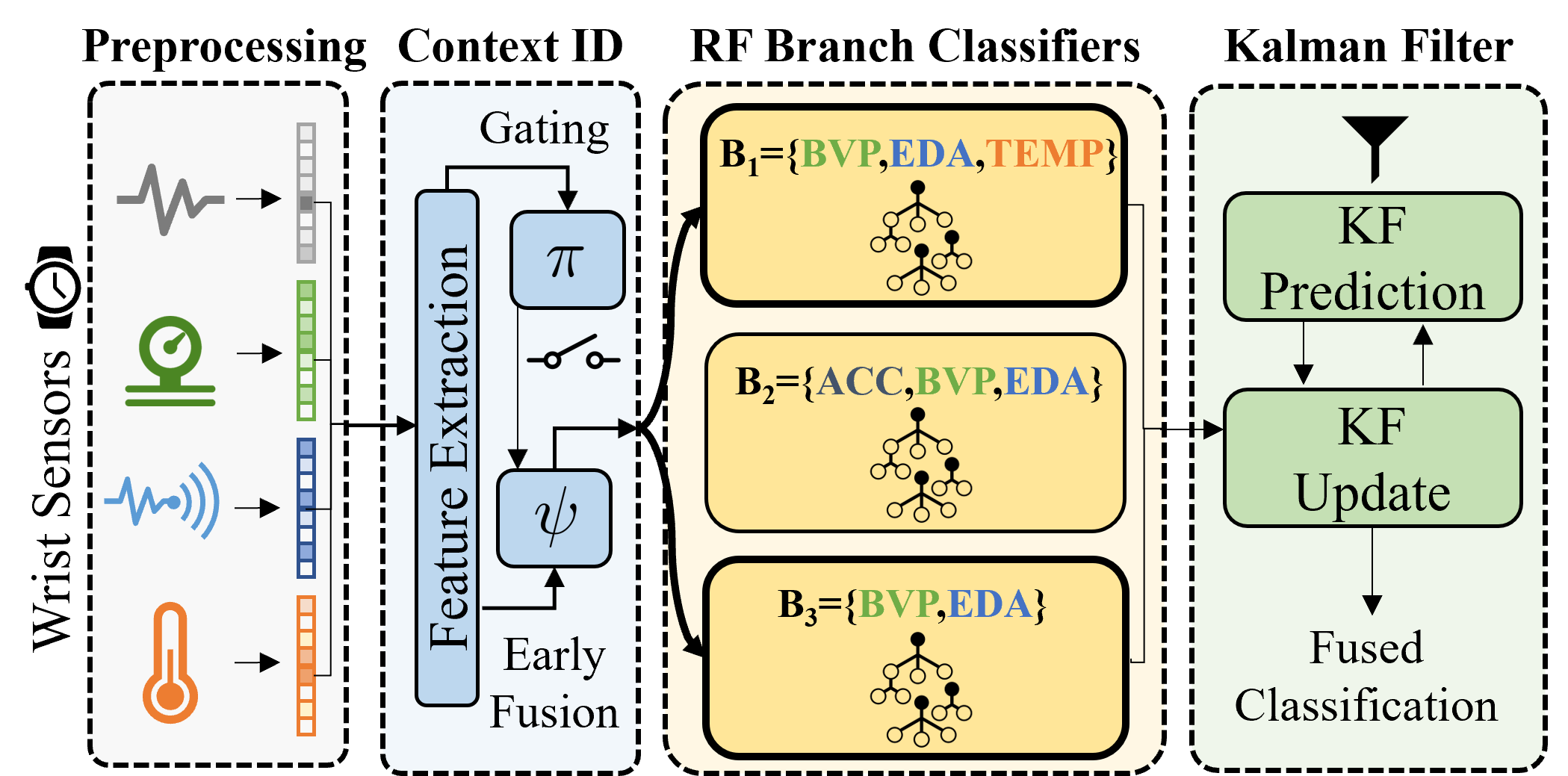}
    \caption{Proposed SELF-CARE Architecture. In this depiction four types of wrist-worn sensors are used, the gating model selects two branches given the context, a Random Forest classifier is used for the branch models, and a Kalman filter is used for the late fusion over the two selected branches.}
    \label{fig:arch}
    \vspace{-5mm}
\end{figure}

In this section we detail SELF-CARE, depicted in Fig. \ref{fig:arch}. Our proposed method performs stress classification given sensor measurements from four wrist sensing modalities: tri-axis accelerometer (ACC), blood volume pulse (BVP), electrodermal activity (EDA), skin temperature (TEMP). 
SELF-CARE uses the four main blocks: (i) preprocessing, (ii) context identification, (iii) branch classifiers, and (iv) late fusion. 


\subsection{Preprocessing Step}
SELF-CARE takes in as inputs data from any number of heterogeneous physiological sensors. Preprocessing is used over the raw, unfiltered sensor data by applying various filters (e.g., band-pass filters or lowpass filters) to the input data to reduce sensor noises and more easily extract important features. The preprocessing performed over each sensing modality follows that performed in \cite{rashid2021feature}. 



\subsection{Context Identification}
\subsubsection{Feature Extraction}
The purpose of the context identification block is to select the branch classifier(s) based on the context of the motion. It first extracts only ACC features as they are directly related to the relative motion of the test subject. These features are then processed by the gating model to select the best performing branch. The feature extraction of the three other modalities takes place after the gating model has selected which branch(es) will be executed. We refer readers to \cite{schmidt2018introducing} for the full list of features per sensor.
\subsubsection{Gating Model ($\pi$)}
The gating model trains a classifier that uses the ACC features as inputs to select one of the available branch classifiers for branches $B_{1}$=\{BVP, EDA, TEMP\}; $B_{2}$=\{ACC, BVP, EDA\}; $B_{3}$=\{BVP, EDA\}. 
A Decision Tree (DT) classifier is used for our gating model, as it is lightweight and adds minimum overhead for our architecture.
Note that, for each round of leave-one-subject-out (LOSO) validation, only training data is used to generate gating labels. Additionally, one, two or all the final classifiers may be selected for final classification depending on the value of $\delta$, detailed next.
\subsubsection{Performance-Energy Trade-off ($\delta$)}
\label{sec:delta}
An important feature of SELF-CARE is its ability to balance constrains between performance and energy. We introduce the term $\delta$ that aids the gating decision in considering this trade-off. The gating model outputs prediction probabilities for the available branches with $\bar{b}$ representing the maximum probability branch. $\delta$ has a range between 0 and 1, representing the range in which non-maximum branches are selected by allowing branches with probabilities greater than $\bar{b}-\delta$ to be also selected. Lower $\delta$ values indicate tighter energy constraints, with $\delta=0$ indicating that only the highest probability branch from the gating classifier is selected, while higher $\delta$ values allow more branches to be selected, with $\delta=1$ indicating that all possible branches are selected. 
For our 3-class (2-class) classification problem we set $\delta=0.4$ ($\delta=0.1$).
\subsubsection{Early Fusion ($\psi$)}
Once the branches are selected after applying $\delta$ on the gating model decision, the features for those branches will be extracted and concatenated together to be passed to the corresponding classifiers (branches). In the example in Fig. \ref{fig:arch}, $B_1$ and $B_3$ are the selected branches from the gating model. The features from BVP, EDA, and TEMP signals are concatenated together using early fusion and fed to the branch classifier for $B_1$, with $B_3$ operating in similar fashion for its sensor modalities.

\subsection{Branch Classifiers}
Next, the corresponding branch classifier(s) is (are) used to perform classification of the segment. 
To train the individual branch classifiers within SELF-CARE we train using different combinations of input sensor data. For our analysis, we use five different early fusion combinations of wrist sensors as input branches - $B_{1}$=\{BVP, EDA, TEMP\}; $B_{2}$=\{ACC, BVP, EDA\}; $B_{3}$=\{BVP, EDA\}; $B_{4}$=\{ACC, BVP\}; $B_{5}$=\{ACC, EDA\}. Each branch is evaluated on five different machine learning classifiers --- Decision Tree (DT), Random Forest (RF), AdaBoost (AB), Linear Discriminant Analysis (LDA), K- Nearest Neighbor (KNN). The classifiers are chosen to ensure a fair comparison with the original WESAD work \cite{schmidt2018introducing}. Additionally, the low complexity of the classifiers makes SELF-CARE suitable for wearable devices. Following the work in \cite{schmidt2018introducing}, we use same configurations for the classifiers.
Out of the 25 (5 branches x 5 classifiers per branch) possible branch classifiers, the branches with the minimum training loss are selected to be used within SELF-CARE. 
Each selected branch outputs a classification prediction to be fused by the late fusion method.


\newcommand{\bad}[0]{\cellcolor{red!10}}
\newcommand{\good}[0]{\cellcolor{green!10}}

\begin{table*}[t]
	\centering
	\caption{Overall Performance Comparison of Related Works using LOSO Validation}
	\label{overall_performance}
	\begin{tabular}{|c||c|c|c||c|c|c||c|c|}
		\hline
		\multirow{2}{*}{\textbf{Modality Used}} &
		\multicolumn{3}{c||}{\textbf{3-Class}} &\multicolumn{3}{c||}{\textbf{2-Class}} & \textbf{Wrist} & \textbf{Wrist} \\
		\cline{2-4}\cline{5-7}
		& \textbf{Best Model} & \textbf{Macro F1} & \textbf{Accuracy} & \textbf{Best Model} & \textbf{Macro F1} & \textbf{Accuracy} & \textbf{Only} & \textbf{Computing} \\
		\hline\hline
		\multicolumn{9}{|c|}{\cellcolor{blue!5}\textbf{Related Works}}\\
		\hline
		All (Wrist+Chest) \cite{schmidt2018introducing} & AB & 68.85 & 79.57 & LDA & 90.74 & 92.28 & \bad No & \bad No \\
		\hline
		{All Wrist \cite{schmidt2018introducing}} & AB & 64.12 & 75.21 & RF & 84.11 & 87.12 & \good Yes & \good Yes \\
		\hline
		{All Wrist + Trans. Chest\cite{samyoun2020stress}} & GAN-RF	& \textbf{74.5} & 81.4 & GAN-RF & 89.7 & 92.1 & \bad No & \bad No \\
		\hline
		{All Wrist \cite{huynh2021stressnas}}  & DNN & - & 83.43 & DNN & - & 93.14 & \good Yes & \bad No \\
		\hline
		{BVP (Wrist) \cite{rashid2021feature}}  & HCNN & 64.15 & 75.21 & HCNN & 86.18 & 88.56 & \good Yes & \bad No \\
		\hline\hline
		\multicolumn{9}{|c|}{\cellcolor{blue!5}\textbf{Selected Branch Classifiers [Ours]}}\\
		\hline
		{\textbf{$B_{1}$=\{BVP, EDA, TEMP\}(Wrist)}} & RF	 & 62.73 & 76.62 & RF & 84.66 & 89.01 & \good Yes & \good Yes \\
		\hline
		{\textbf{$B_{2}$=\{ACC, BVP, EDA\}(Wrist)}} & RF & 62.88 & 77.71 & RF & 85.08 & 88.76 & \good Yes & \good Yes \\
		\hline
		{\textbf{$B_{3}$=\{BVP, EDA\}(Wrist)}} & RF & 61.02 & 73.96 & RF & 86.37 & 89.33 & \good Yes & \good Yes \\
		
		\hline\hline
		\multicolumn{9}{|c|}{\cellcolor{blue!5}\textbf{Traditional Late Fusion [Ours]}}\\
		\hline
		\textbf{Soft-voting ($B_{1}$, $B_{2}$, $B_{3}$)} & RF	& 63.75	& 78.79 & RF & 87.09 & 90.00 & \good Yes & \good Yes \\
		\hline
		\textbf{Hard-voting ($B_{1}$, $B_{2}$, $B_{3}$)}  & RF & 64.02	& 78.70 & RF & 87.17 & 89.89 & \good Yes & \good Yes \\
		\hline\hline
		\multicolumn{9}{|c|}{\cellcolor{blue!5}\textbf{SELF-CARE [Ours]}}\\
		\hline
		\textbf{Kalman ($B_{1}$, $B_{2}$, $B_{3}$)} & RF & 71.97 & \textbf{86.34} & RF & \textbf{92.93} & \textbf{94.12} & \good Yes & \good Yes \\
		\hline
		
	\end{tabular}
	\vspace{-3ex}
\end{table*}

\subsection{Late Fusion Method}
\label{sec:lfb}
Here we present our \textit{Kalman filter-based} method for classification over an ensemble of classifiers, although we claim that any applicable late fusion method is supported within SELF-CARE.
In the context of our problem, we consider a Kalman filter approach towards the multi-class classification problem like in \cite{pakrashi2019kalman}, however, we additionally model the temporal dynamics in the stress classification problem for each sample. The unknown state our filter is attempting to estimate is the probability of each class during each segment. Thus, we define this $\mathbf{x}$ as a $c$ dimensional vector of estimated class probabilities. Additionally, the predictions from each separate classifier are the measurements $\mathbf{z}$, which are processed sequentially per time step. For the 3-class (2-class) problem, we initialize $\mathbf{x}_0 = [0.8,0.1,0.1]^\top$ ($\mathbf{x}_0 = [0.8,0.2]^\top$) with estimation error covariance matrix $\mathbf{P}_0 = 0.01\cdot\mathbf{I}_{3x3}$ ($\mathbf{P}_0 = 0.01\cdot\mathbf{I}_{2x2}$). The state transition matrix and measurement matrix are identity matrices for the respective problems. The process noise for both problems is modeled as a discrete time white noise with variance set at 5e-4. The measurement noise is modeled as a function of each measurement to allow the filter to adjust the confidence of the measurements according to each reported class probability: $\mathbf{R} = ((\mathbf{1}-\mathbf{z})\cdot2\cdot\mathbf{I}_{3x3})^2$ ($\mathbf{R} = ((\mathbf{1}-\mathbf{z})/2\cdot\mathbf{I}_{2x2})^2$).
Lastly, a tunable threshold technique was used to process the measurements which involved (i) an $\epsilon$ parameter to select measurements which had a maximum predicted probability above the threshold and (ii) a $\gamma$ factor to scale the measurements to account for the imbalanced class distribution in the dataset. This thresholding process allows for the filter to weight each measurement it receives with a differing degree of noise while also attempting to resolve issues that arise from imbalanced datasets. For the 3-class (2-class) problem, we set $\epsilon = 0.4$ ($\epsilon = 0.7$) and $\gamma = [.278,1,1]^\top $ ($\gamma = [.667,1.1]^\top$). During 3-class classification, the prediction probabilities are generally lower as they are distributed across an additional class when compared to 2-class classification, thus calling for a lower $\epsilon$ threshold.
To validate our Kalman-filter based method, we benchmark its performance against commonly used voting mechanisms for late fusion: \textit{hard-voting} and \textit{soft-voting} \cite{oviatt2018handbook}. The method of hard-voting assigns the final class based on the class most commonly voted by each classifier, whereas soft-voting selects the class with the highest average value across all the classifiers. 

\section{Experimental Analysis}
\label{sec:results}

\subsection{Dataset Evaluation Metrics}
SELF-CARE is validated on the publicly available WESAD dataset \cite{schmidt2018introducing}. The dataset contains data for a total of 15 subjects from both chest (RespiBAN) and wrist (Empatica E4) worn sensors. Our work focuses on stress detection using only wrist-based data, as we use the following sensors from the Empatica E4: ACC BVP, EDA, TEMP. 
The dataset has three types of classes related to emotional states, namely --- baseline (neutral), amusement, and stress. For the 2-class problem, baseline and amusement are considered as the non-stress class. The filtered signals are segmented by a window of 60 seconds of data with a sliding length of 5 seconds following \cite{samyoun2020stress}. This gives a total of 6458 segments for each signal across all subjects of the WESAD dataset. The WESAD dataset is highly imbalanced in terms of the number of segments per class. For this reason, F1 score is also used along with accuracy to measure the classification performance. To ensure a fair comparison with other works, we use the macro F1 score.

\subsection{Experimental Results}
This section presents the performance of SELF-CARE for stress detection in 3-class and 2-class classification. We also demonstrate the energy efficiency of our approach in a ultra-low-power 32-bit microcontroller EFM32 Giant Gecko (EFM32GG-STK3700A) \cite{EFMGG} representing a wearable device operating on the edge. The microcontroller has an ARM Cortex-M3 processor with a maximum clock rate of 48 MHz. It has 128 KB of RAM and 1 MB of flash memory. 


\subsubsection{Performance Evaluation}
\label{performance_evaluation}

Table \ref{overall_performance} shows the overall performance comparison of the related works against our proposed method. Authors in \cite{schmidt2018introducing} explored different combinations of chest and wrist sensors across a variety of models. The results for three deep learning methods are also shown \cite{samyoun2020stress,huynh2021stressnas,rashid2021feature}. 
For our three selected branch classifiers, the soft- and hard-voting methods are applied, showing performance improvements compared to the individual branch classifiers for both 3-class and 2-class classification. Lastly, SELF-CARE using Kalman filter-based late fusion further improves the performance for 3-class and 2-class classification compared to these traditional late fusion methods.
Despite using only wrist signals, SELF-CARE outperforms all other state-of-the-art works that use either wrist, chest, or both sensors for 3-class and 2-class classification. Only \cite{samyoun2020stress} achieves a better macro F1 score than SELF-CARE for 3-class classification. However, they use both wrist and translated chest features, and employ a computationally expensive GAN model, which is not suitable for wrist computing.

\subsubsection{Energy Evaluation}

\begin{figure}[t]
		\centering
		\includegraphics[trim={9cm 15.74cm 11.5cm 5.7cm},clip, width=\linewidth]{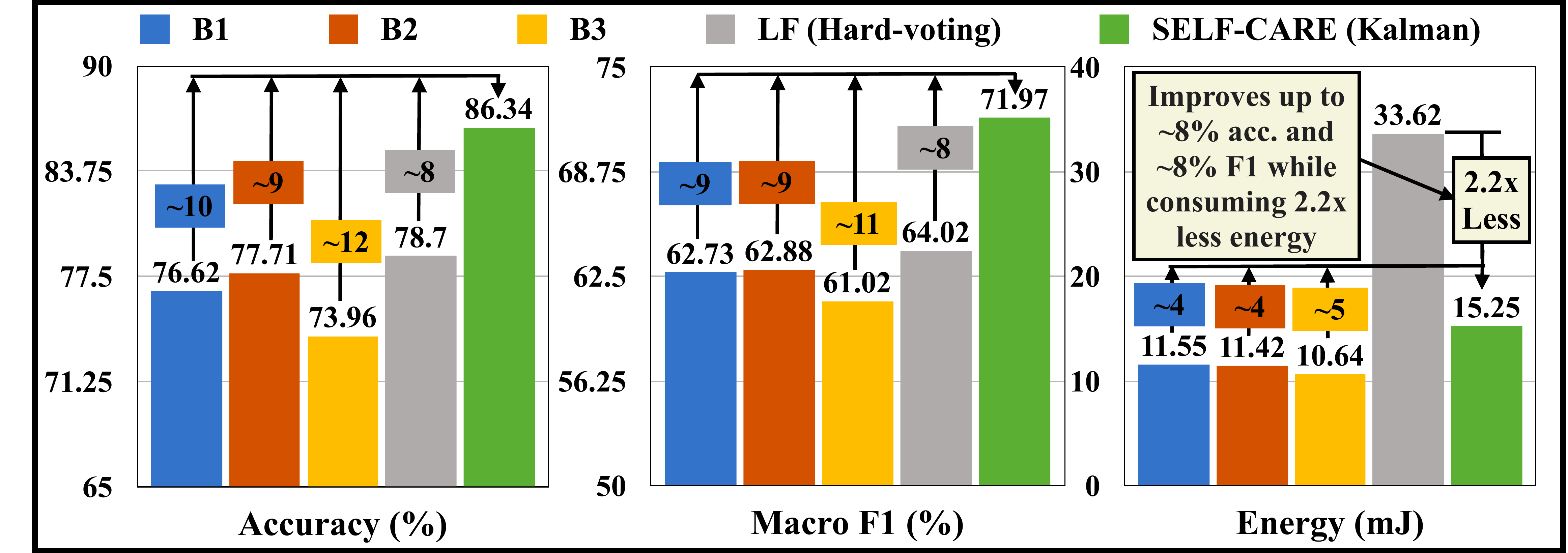}
		\caption{Performance and Energy for 3-Class Classification}
		\vspace{-3mm}
		\label{3_class_benchmark}
\end{figure}

\begin{figure}[t]
		\centering
		\includegraphics[trim={9cm 15.74cm 11.5cm 5.7cm},clip, width=\linewidth]{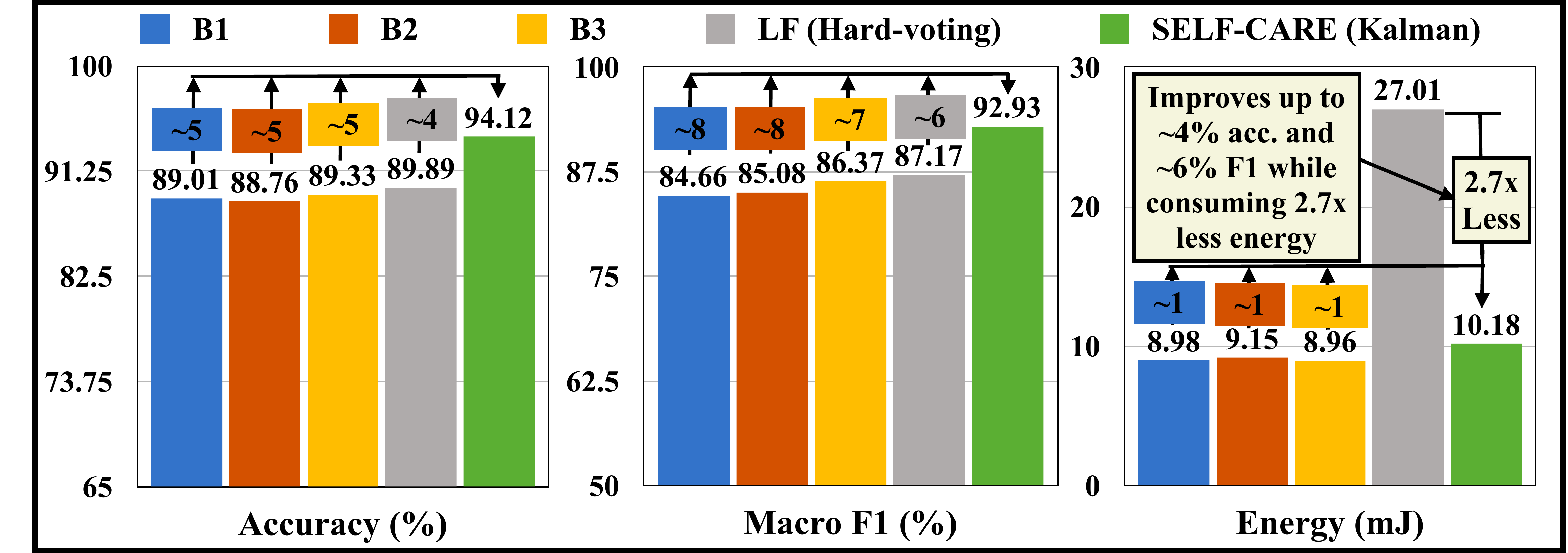}
		\caption{Benchmarking on 2-Class Classification}
		\vspace{-5mm}
		\label{2_class_benchmark}
\end{figure}

As shown in Table \ref{overall_performance}, traditional late fusion improves the performance compared to individual branch classifiers. However, it is not energy-efficient, as multiple classifiers need to be used simultaneously --- unlike SELF-CARE that minimizes the number of classifiers selected for a given segment.
We benchmark SELF-CARE with hard-voting late fusion, which is relatively more energy-efficient than soft-voting and shows similar performance to soft-voting. 
As shown in Fig. \ref{3_class_benchmark} for 3-class classification, SELF-CARE with $\delta=0.4$ improves up to $\sim$8\% accuracy and $\sim$8\% F-1 score, while being $\sim$2.2$\times$ energy-efficient compared to hard-voting. Similarly, for 2-class classification (Fig. \ref{2_class_benchmark}), SELF-CARE with $\delta=0.1$ outperforms hard-voting by up to $\sim$4\% accuracy and $\sim$6\% F-1 score while being $\sim$2.7$\times$ energy-efficient. The higher energy efficiency for 2-class can be partially attributed to the lower $\delta=0.1$, which reduces the use of multiple branches compared to $\delta=0.4$ for 3-class.
The higher $\delta$ for 3-class is chosen to prioritize performance over energy, as the 3-class problem is inherently more challenging than 2-class.


\section{Conclusion}
\label{sec:con}
In this paper we proposed SELF-CARE, a selective sensor fusion approach that uses context-aware, energy-efficient edge computing to perform stress detection. SELF-CARE models context as the motion of a subject and performs an intelligent gating mechanism to select which sensor fusion schema to use given a certain input. 
To the best of our knowledge, SELF-CARE achieves state-of-the-art performance on the WESAD dataset in terms of 3-class classification (\textbf{86.34\%}) and 2-class classification (\textbf{94.12\%}) in approaches that use LOSO validation. Furthermore, SELF-CARE achieves upto \textbf{2.2$\times$} (3-class) and \textbf{2.7$\times$} (2-class) energy efficiency with respect to comparable late fusion methods. 

\vspace{-1mm}
\section*{Acknowledgment}
This work was partially supported by the National Science Foundation (NSF) under awards CMMI-1739503 and CCF-2140154. Any opinions, findings, conclusions, or recommendations expressed in this paper are those of the authors and do not necessarily reflect the views of the funding agencies.
\vspace{-1mm}

\bibliographystyle{IEEEtran}
\bibliography{bibliography}

\end{document}